\documentclass[twocolumn,journal]{IEEEtran}
\usepackage{ifpdf}
\usepackage{url}
\usepackage{amsmath}
\usepackage{comment}
\usepackage{color}
\usepackage{enumerate}
\usepackage{amssymb}
\usepackage{bbm}
\usepackage{mleftright}
\usepackage{gensymb}
\usepackage{booktabs}
\usepackage{diagbox}
\usepackage{hhline}
\usepackage{balance}
\usepackage{multirow}
\usepackage{textcomp}
\usepackage{subcaption}
\usepackage{textcomp}
\usepackage{xcolor}
\usepackage{csquotes}

\usepackage{amsthm}

\DeclareMathAlphabet{\pazocal}{OMS}{zplm}{m}{n}

\ifCLASSINFOpdf
  \usepackage[pdftex]{graphicx}
  \DeclareGraphicsExtensions{.pdf,.png,.bmp}
\else
  \usepackage[dvips]{graphicx}
  \DeclareGraphicsExtensions{.eps}
\fi

\begin{document}

\colorlet{MajorRevision}{black!}


\title{Integrated Access and Backhaul in Millimeter-Wave Cellular: Benefits and Challenges \vspace{-2mm}}


\author{Yekaterina~Sadovaya, Dmitri~Moltchanov, \\Wei Mao, Oner Orhan, Shu-ping~Yeh, Hosein~Nikopour, Shilpa Talwar, and Sergey~Andreev
\thanks{Y. Sadovaya, D. Moltchanov, and S. Andreev are with Tampere University, Finland. Email:~{firstname.lastname}@tuni.fi}
\thanks{W. Mao, O. Orhan, S.-p. Yeh, H. Nikopour, and S. Talwar are with Intel Corporation, Santa Clara, CA, USA. Email:~{firstname.lastname}@intel.com}
\vspace{-0mm}
}

\maketitle

\begin{abstract} 
The recently proposed NR-ready integrated access and backhaul (IAB) architecture promises to bring a cost-efficient deployment solution for both coverage extension and capacity boosting in future 5G/5G+ systems. While its impact on the coverage extension was thoughtfully addressed in the literature, the effect of advanced functionalities such as multi-hop, multi-connectivity, and multi-beam operations on the throughput remains unclear. \textcolor{MajorRevision}{We review and characterize the system-level impact of these capabilities on the performance of self-backhauled IAB systems operating in half-duplex mode and utilizing millimeter-wave (mmWave) technology across both access and backhaul. Our results indicate that the throughput gain of multi-hopping and multi-beaming is significant even without multi-connectivity operation. Another important learning is that in all-mmWave systems with link blockage, multi-connectivity with link switching allows achieving self-load balancing. Finally, we outline future research directions.}

\end{abstract}


\section{Introduction}




While the first wave of 5G New Radio (NR) deployments utilizing microwave bands is already underway, the attention of operators and vendors is now set on millimeter-wave (mmWave) band NR that allows benefiting from larger bandwidth. \textcolor{MajorRevision}{According to \textcolor{MajorRevision}{the 3rd Generation Partnership Project (3GPP)} Release 17, frequencies up to 71 GHz are supported.} However, severe path loss along with link blockage require extremely dense deployments to provide ubiquitous \textcolor{MajorRevision}{coverage \cite{alammouri2020escaping, Cudak2021}}. Exploring the ways to reduce capital expenditures when deploying mmWave 5G NR systems, 3GPP has recently proposed integrated access and backhaul (IAB) architecture \cite{madapatha2020integrated}. 


By utilizing relays, named IAB nodes, 3GPP incorporates inherently multi-hop architecture into future 5G/5G+ cellular system design with backhaul links connecting IAB nodes to each other and to the donor gNB (DgNB). The user equipment (UE) that is outside the coverage of the DgNB may associate with the nearest available IAB node, thus benefiting from coverage extension provided by the emerging IAB architecture \cite{polese2020integrated}. The performance of an IAB network with respect to the coverage extension has been extensively investigated in the literature \cite{lukowa2020}. \textcolor{MajorRevision}{Resource partitioning is addressed in \cite{Pagin2022} while topology formation problem is solved in \cite{huang2021bayesian, simsek2020iab}.}






The throughput of 3GPP IAB multi-hop systems is limited by the constraints of modern wireless principles such as the half-duplex operation \cite{saha2019millimeter}. To improve the capacity while utilizing the IAB architecture, 3GPP included the support of multi-connectivity and multi-beam functionalities as well as dynamic slot formatting. These capabilities are aimed at throughput boosting and are crucial for all-mmWave IAB deployments, where the mmWave band is utilized for both access and backhaul \cite{8514996}.  


The multi-beam operation is expected to enhance the amount of available radio resources at the expense of increased interference while multi-connectivity may potentially allow for higher connection reliability and enhanced data rates at the air interface. However, the impact of all these mechanisms depends on many factors including the scenario of interest, constraints imposed by multi-hop operation and half-duplex IAB radio design, specific implementations of these functionalities, as well as on the utilized resource allocation (RA) scheme. Therefore, the ultimate effect of these mechanisms on the system throughput in all-mmWave 3GPP IAB architecture still remains unclear.



In this article, we study the effects of advanced IAB functionalities primarily on per-user throughput in all-mmWave 5G NR deployments. By gradually adding multi-hopping, multi-beaming, and multi-connectivity, we isolate and thoroughly characterize the impact of each individual mechanism as well as their joint utilization under static and dynamic slot formatting schemes. Then, we discuss several challenges for implementing these features in real-world systems. The results help to outline a set of recommendations on the use of these capabilities in practical all-mmWave 3GPP IAB deployments. 




The rest of this text is organized as follows. First, in Section \ref{sect:arch}, we review the state-of-the-art IAB architecture and introduce the considered advanced functions with their various design options. The numerical results and their interpretation are provided in Section \ref{sect:results}. Section \ref{sect:challenges} highlights the main challenges associated with the implementation of the advanced functionalities for future deployments of the IAB technology. Conclusions are drawn in the last section.

\section{IAB Architecture and Enhancements}\label{sect:arch}

In this section, we start with the overall review of the IAB system architecture. Then, we introduce the considered functionalities individually.

\subsection{3GPP IAB Architecture for In-Band Backhauling}

The architecture of the NR-ready IAB system is \textcolor{MajorRevision}{based on the CU/DU split, which was proposed in 3GPP TR 38.874 Release 16. It is facilitated by new entities named IAB nodes.} \textcolor{MajorRevision}{The details on higher-layer protocols and architecture are later specified in TS 38.401, while radio transmission and reception for IAB are documented in TS 38.174. The current work in Release 17 related to IAB enhancements includes improved topology robustness, resource multiplexing, and network management. The possible extensions of IAB in Release 18 may include the self-interference mitigation methods to improve the full-duplex operational mode, enhanced mobility support, and reduced multi-hop latency.} 

The gNB is connected to the core network via the NG interface and provides the user plane (UP) and control-plane (CP) protocols termination toward UEs. The gNB can be a single logical node or it can comprise a central unit (CU) and a distributed unit (DU) connected with each other via the F1 interface. Each IAB node handles MT and DU, where the MT function is responsible for communication with a parent node while the DU function arbitrates communication with a child. According to 3GPP TR 38.874, in-band operations are inherently limited by the half-duplex constraint. In this case, an IAB node cannot receive and transmit simultaneously. 

The concept of IAB assumes the reuse of the 5G access links for backhauling. It utilizes the functions of mobile-termination (MT), gNB-DUs, gNB-CU, user plane function (UPF), mobility management function (AMF), session management function (SMF), and the corresponding interfaces, i.e. NR Uu, F1, NG, X2, and N4. The IAB node connects to its parent using the MT functionalities over the NR Uu link. \textcolor{MajorRevision}{The protocol stack for the CU/DU split architecture is specified in TS 38.401.}

\begin{figure}
    \centering
    \includegraphics[width=0.95\columnwidth]{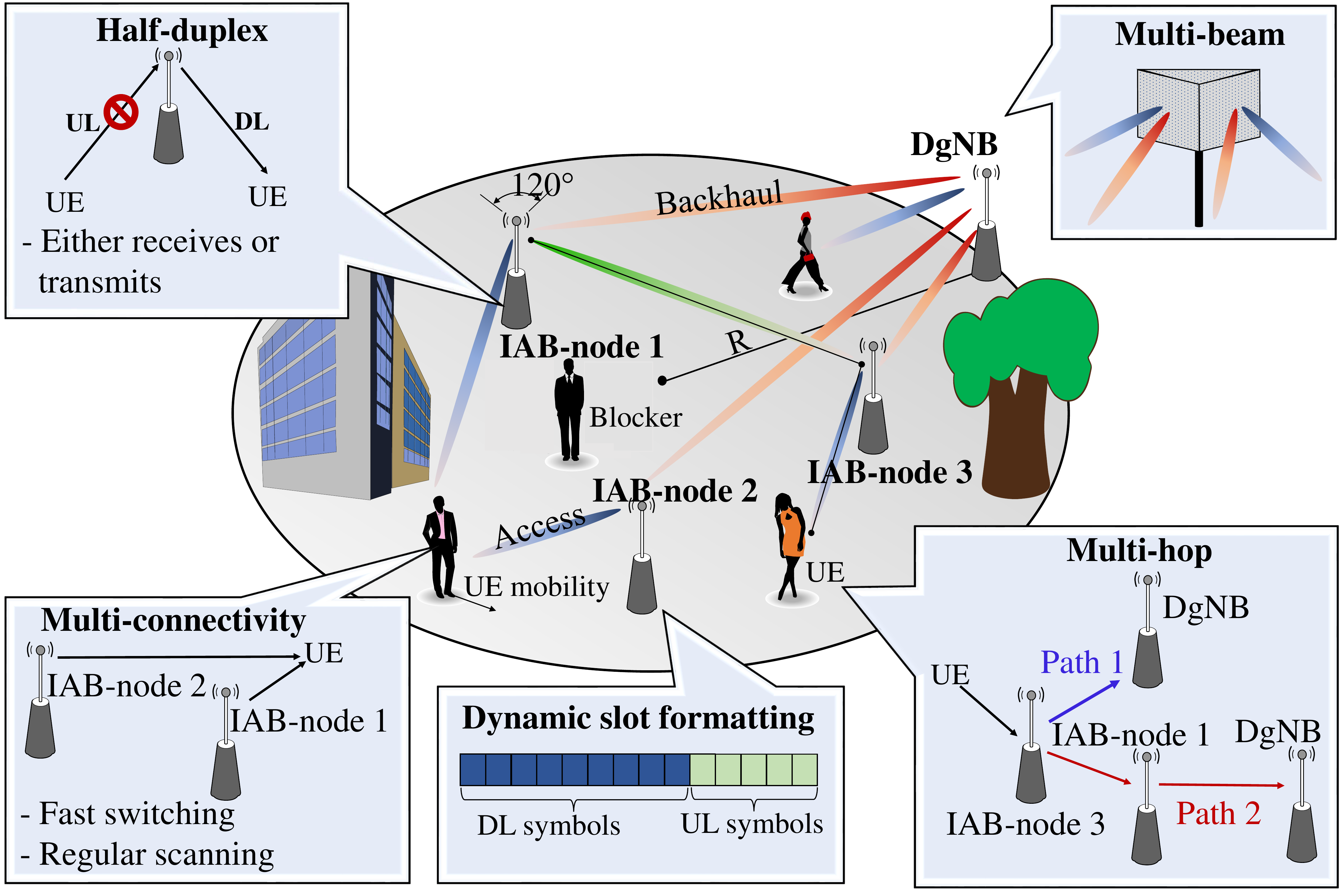}
    \caption{IAB deployment of interest.}
    \label{fig:deployment}
    \vspace{-4mm}
\end{figure}

\subsection{Multi-Hop Operation}

IAB nodes enable multi-hop backhauling, which allows for flexible coverage extension in 5G NR deployments. An example of the IAB architecture with multi-hop functionality is offered in Fig. \ref{fig:deployment}. In this setup, each IAB node and DgNB holds zero or more child nodes, which are located below it in the tree. A node that has a child is called a parent node. 

\textcolor{MajorRevision}{To establish an IAB network, the so-called integration procedure is accomplished. According to TS 38.401, a parent node is discovered initially. After that, the IAB node requests an RRC connection with the CU via the parent node while the backhaul link is created via the RLC. IAB-specific features include backhaul adaptation protocol (BAP), which is defined in TS 38.340}. It is employed at the backhaul links to enable efficient multi-hop forwarding.

The access and backhaul links at the IAB nodes can be multiplexed using time-division (TDM), frequency division (FDM), or space division multiplexing (SDM). \textcolor{MajorRevision}{However, as it is stated in 3GPP TS 38.174, IAB is especially beneficial at mmWave spectrum; hence, TDM is the most common approach due to the large available bandwidth.}
Following 3GPP TR 38.874, RLC between IAB nodes can be hop-by-hop or end-to-end. End-to-end automatic repeat request (ARQ) can be beneficial as packets do not traverse through all RLC states at intermediate IAB nodes. On the other hand, hop-by-hop ARQ guarantees more efficient retransmissions.

\textcolor{MajorRevision}{According to TR 38.874, directed acyclic graph (DAG) and tree multi-hop topologies are supported. The procedure of the intra-CU backhaul radio link failure (RLF) recovery is later described in TS 38.401, which is required to be performed by IAB nodes to switch from one parent to another parent node under the same IAB donor CU.}



\subsection{Multi-Beam Functionality}


Another advanced system feature is multi-beam functionality that can be utilized in IAB deployments at both DgNB and IAB nodes. Multi-beam communications imply simultaneous operation of independent directional beams, which enables efficient frequency reuse and significantly higher system capacity. However, the transmit power of an individual beam is reduced as compared to a single-beam scenario due to the fact that it is split among the beams. 3GPP provides more details regarding beam management in Release 14 TR 38.912, while alternative strategies are proposed in, e.g., \cite{li2020beam}.

The multi-beam capabilities can be utilized at both DgNB and IAB nodes. However, this functionality may significantly compromise the cost-efficiency of practical NR deployments as it requires digital or hybrid beamforming \cite{palacios2017tracking}. Furthermore, it may also negatively affect the system performance as the transmit power is divided among the beams potentially reducing the resulting coverage and capacity. 

\subsection{Multi-Connectivity Capabilities}



Multi-connectivity improves network reliability via simultaneous support of several links from source to destination. The maximum number of links that can be utilized simultaneously is named the degree of multi-connectivity. Dual-connectivity is ratified in 3GPP TS 37.340, where it implies that a UE utilizes radio resources of two eNB/gNB within the same band. In 5G NR, the dual connectivity notion is generalized, i.e., the UE may exploit the resources provided by E-UTRA access and NR access simultaneously. 

In the context of IAB networks, multi-connectivity can be implemented in different ways. For example, an additional connection might serve as a backup link or they can be utilized simultaneously. Potentially, other solutions are also feasible and they are addressed in what follows. In our deployment, a given UE utilizes resources of several nodes simultaneously.





By utilizing multi-connectivity, advanced functionalities to combat link blockage can be employed. This functionality refers to fast switching (FS), i.e., changing the association point if the current one becomes unavailable. In this setup, UE does not perform re-switching to the initial state even if the blockage period on that link has expired. The improved version of FS corresponds to the situation where re-switching is allowed even when the links are not blocked. Moreover, regular scanning is utilized for a continuous update of links with the highest RSRP.



\subsection{Dynamic Slot Formatting}

\begin{figure}[!t]
    \centering
    \includegraphics[width=0.9\columnwidth]{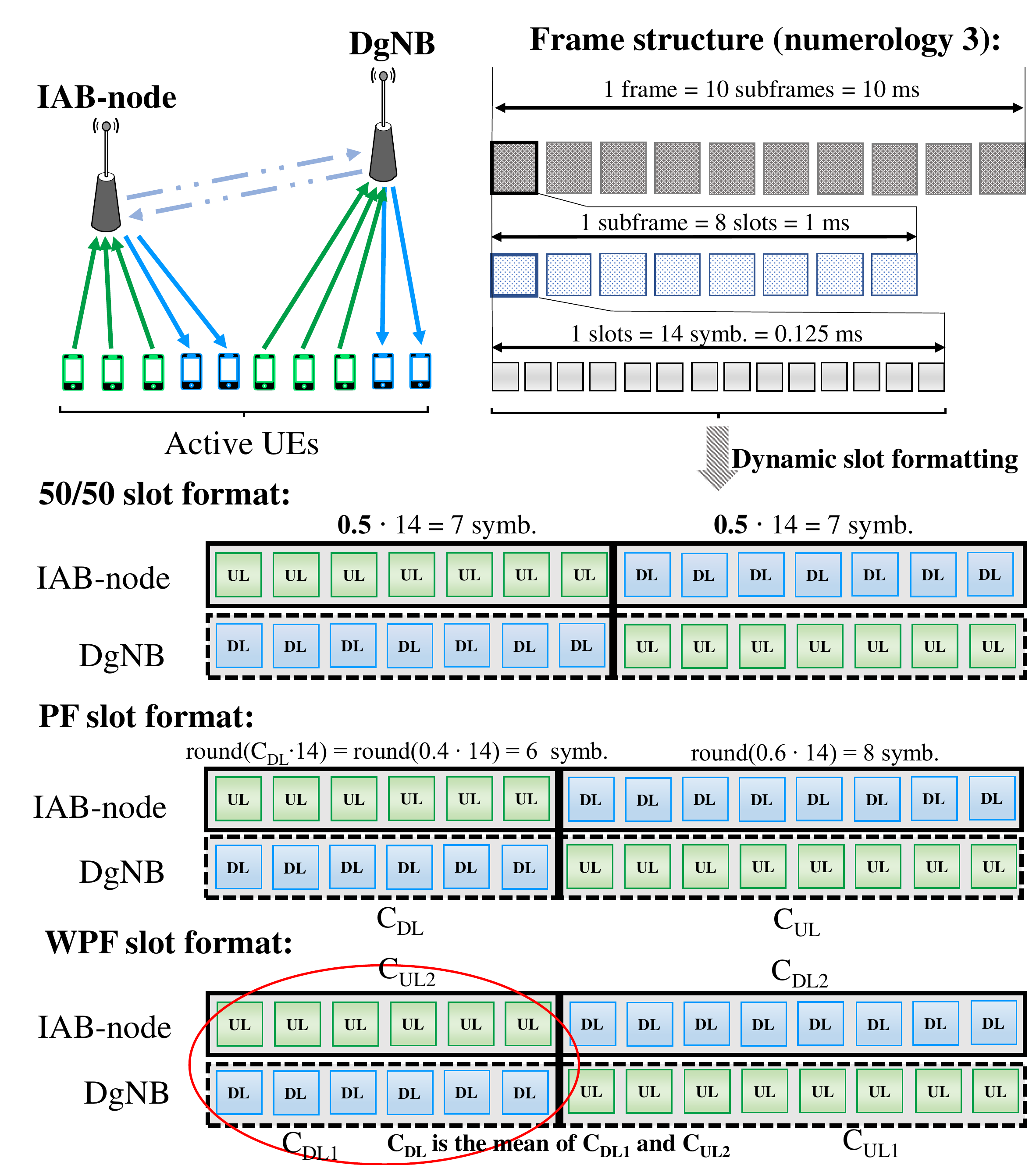}
    \caption{Considered slot formats.}
    \label{fig:ra}
    \vspace{-4mm}
\end{figure}

5G NR offers 6 different waveform configurations, which are known as numerologies. Depending on the numerology, the symbol length and the number of slots within the frame can be controlled to satisfy various throughput and latency requirements. The general frame structure for, e.g., the third numerology is provided in Fig. \ref{fig:ra}. 

Slot formatting for 5G NR systems was introduced in 3GPP TS 38.213. It indicates how each OFDM symbol in a single slot can be utilized. This allows making scheduling more adjustable as compared to LTE. The NR specifications offer 61 predefined symbol combinations, which can be assigned while designing the network. In addition, dynamic slot configuration can be enabled, which is essential for capacity improvement. 

We consider three different methods of slot division shown in Fig. \ref{fig:ra}. The first one is a baseline, where the same amount of resources is allocated for \textcolor{MajorRevision}{uplink (UL)} and \textcolor{MajorRevision}{downlink (DL)} directions. 
On the one hand, the implementation of this division is straightforward. However, asymmetric load in the UL and DL directions is not accounted for. To implement proportional fair (PF) slot formatting, the number of active UEs in UL and DL should be reported. UE is considered active if it has buffered traffic in the UL or DL directions. After this information is provided, the slot division coefficient in the DL can be computed as a fraction of the number of active UEs in the DL to the total number of active UEs in the UL and DL. The slot division coefficient in the UL is calculated similarly. 

The weighted PF (WPF) approach aims to improve the PF method by enhancing fairness. For example, $C_{UL_2}$ in Fig. \ref{fig:ra} is computed as a fraction of the number of active UEs in the UL connected to the IAB-node to the total number of active UEs in the UL and DL connected to this IAB-node. Then, $C_{UL_2}$ and $C_{DL_2}$ (see Fig. \ref{fig:ra}) are averaged to account for asymmetry in the traffic demands aiming to equalize the slot weights in logical directions. 

\begin{table}[t]
\vspace{-0mm}
		\caption{Parameters utilized in numerical assessment.}
		\label{tab:parameters}
		\begin{center}
			\begin{tabular}{p{0.58\columnwidth}p{0.32\columnwidth}}
				\hline
				\textbf{Parameter}                & \textbf{Value}   \\
				\hline\hline
				Carrier frequency         & 30 GHz\\
				\hline
				Bandwidth         & 400 MHz\\
				\hline
				Number of UEs & 60 \\
				\hline
				Cell radius & 500m \\
				\hline
				Tx power of DgNB & 40 dBm \\
				\hline
				Tx power of IAB node & 33 dBm \\
				\hline
				Tx power of UE & 23 dBm \\
				\hline
				No. of IAB nodes & 3, \textcolor{MajorRevision}{7} \\
				\hline
				\textcolor{MajorRevision}{No. of DgNBs} & \textcolor{MajorRevision}{1} \\
				\hline
				Noise figure of DgNB and IAB node & 7 dB \\
				\hline
				Noise figure of UE & 13 dB \\
				\hline
				Power spectral density of noise & -173.93 dBm/Hz\\
				\hline
				Antenna array of UE & 4x4 \\
				\hline
				Antenna array of DgNB and IAB node & 16x16 \\
				\hline
				Velocity of UE & 3 km/h \\
				\hline
				Height of DgNB & 25 m \\
				\hline
				Height of IAB node & 10 m \\
				\hline
				Height of UE & 1.5 m \\
				\hline
				Height of blocker & 1.5 m \\
				\hline
				Radius of blocker & 0.2 m \\
				\hline
				Degree of multi-connectivity & 2 \\
				\hline
				Scheduler & Round-robin \\
				\hline
				File Size & 2 Mbytes \\
				\hline
			\end{tabular}
		\end{center}
		\vspace{-4mm}
	\end{table}

\section{Performance Evaluation Results}\label{sect:results}

In this section, we report the results of our evaluation campaign by focusing on the impact of each of the considered functionalities in detail. 


\subsection{Simulation Scenario}


Our scenario \textcolor{MajorRevision}{consists of one cell while} the choice of parameters is based on the 3GPP recommendations for the IAB system evaluation, which are defined in TR 38.874. The parameters utilized for the numerical assessment are provided in Table \ref{tab:parameters}. The scenario is illustrated in Fig. \ref{fig:deployment}. The DgNB is placed at the cell edge while the IAB nodes and UEs are dropped randomly. Moreover, the UEs are mobile \textcolor{MajorRevision}{while link blockage is modeled according to \cite{gapeyenko2017temporal}. The blockage model is based on the alternating process of blocked and unblocked states, which represents the blockage caused by a human body. The initial blockage probability and the duration of blocked and unblocked intervals depend on the height, radius, and density of blockers. The duration of each interval also depends on the average velocity of blockers, which is assumed to be the same as UE velocity (see Table \ref{tab:parameters}). More details can be found in \cite{gapeyenko2017temporal}. Co-channel interference is taken into account using the margin, which is obtained via additional system-level modeling. For the multi-beam case, interference between the beams is suppressed using diversity techniques \cite{sadovaya2021self}.} 

By following the 3GPP recommendations, the traffic of each UE in the UL and DL is assumed to follow FTP model 3. As per the 3GPP guidelines, the channel is modeled according to TR 38.901 with large-scale and small-scale parameters following 3D UMi and UMa deployments depending on the antenna heights (see Table \ref{tab:parameters}). \textcolor{MajorRevision}{The directional antenna patterns are simulated following the TR 38.901 guideline, which implies that the total gain of an array is obtained as the superposition of its elements. The procedure of pattern generation is described in detail in TR 37.840. }

In multi-hop scenarios, we consider two topology formation strategies. The first one corresponds to the situation where the path is selected based on the minimum number of hops between UE and the DgNB. The second strategy refers to the case where the path is selected based on the maximum reference signal received power (RSRP).  \textcolor{MajorRevision}{It implies that the backhaul route is chosen according to the maximum RSRP value of the worst link in all the available routes.} The first method reflects the situation where all the UEs attempt to connect to the DgNB if it is available. The scheme based on the maximum RSRP criterion has been proposed in several 3GPP contributions, e.g., R1-1808692 and R1-1811514.

\textcolor{MajorRevision}{We also consider different multi-beam options to quantify their effect on the UE throughput. First, a fully multi-beam scenario is considered, where all IAB nodes and the DgNB have separate beams for backhauling and an additional access beam. Further, a scenario where only the DgNB has multi-beam functionality is addressed. Finally, an all-single beam scenario is considered, where all IAB nodes and the DgNB utilize the single beam configuration. In all the simulations, the statistical data are obtained in the steady-state regime.}

\subsection{Multi-Hopping and Dynamic Slot Formatting}

\begin{figure}[!b]
    \vspace{-4mm}
    \centering
    \includegraphics[width=0.9\columnwidth]{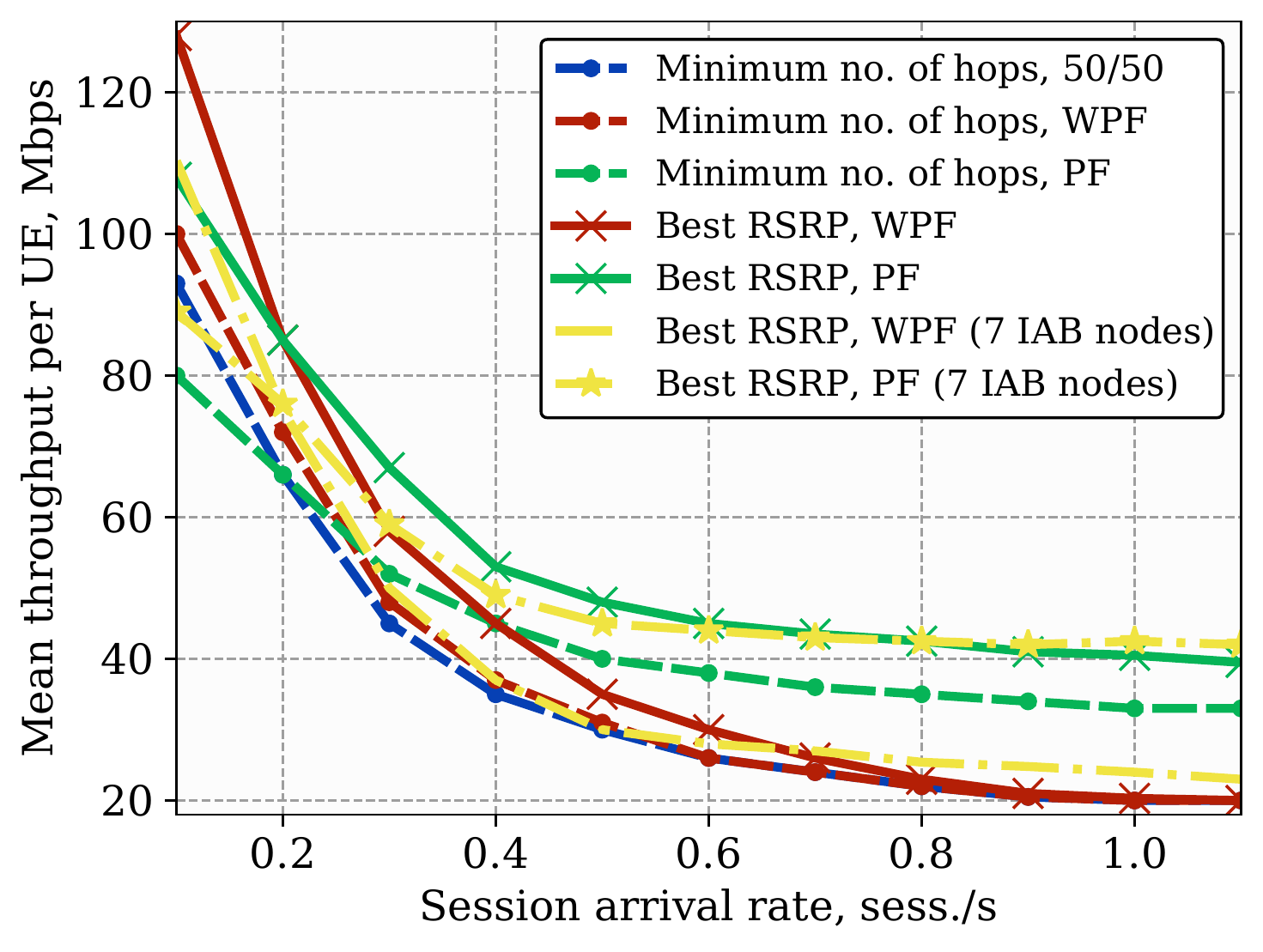}
    \caption{\textcolor{MajorRevision}{Mean throughput per UE as a function of session intensity for multi-hopping schemes based on the minimum number of hops and the highest RSRP value.}}
    \label{fig:multi_hop}
    \vspace{-0mm}
\end{figure}

We start with the impact of multi-hopping for a tree topology as illustrated in Fig. \ref{fig:multi_hop}, where the mean throughput per UE is given as a function of the DL session arrival intensity for a fixed UL session intensity of $0.2$ sessions per second. \textcolor{MajorRevision}{In this configuration, multi-hopping schemes are utilized with the single-beam operation and no multi-connectivity support.} The session intensity defines the probability of arrival per time unit, in the UL it is fixed to $0.2$ as it keeps the system from overloading. As can be seen, the association scheme based on the minimum number of hops yields $15-20$\% lower mean throughput. The explanation is twofold. First, static slot formatting in the UL and DL directions is not optimal under dynamic traffic conditions. On top of this, several UEs are forced to connect to the DgNB despite poor channel conditions. \textcolor{MajorRevision}{Increasing the number of IAB nodes from 3 to 7 at first decreases the throughput because of the larger delay. However, for higher values of session intensities, the same schemes with 7 IAB nodes provide larger throughput than with 3 IAB nodes due to the load balancing effect.}

Understanding the obtained results further, one may notice that WPF converges to the 50/50 in the considered scenario, i.e., the slot division coefficient equals 0.5 most of the time. Conversely different behavior is exhibited by the PF approach. This is because the PF scheme allocates more resources to the more loaded direction. In addition, this effect lowers the backlog in the overloaded direction.
Moreover, the use of dynamic slot formatting in multi-hop regime improves the UE throughput by $10 - 30$ \% even when other advanced capabilities including multi-connectivity and multi-beaming are not utilized. In practice, it implies that cost-efficient 3GPP IAB solutions using simple single-beam antenna arrays and CUs not supporting multi-connectivity capabilities may still greatly benefit from optimized dynamic slot formatting.



\subsection{Multi-Connectivity with Advanced Link Selection}

\begin{figure}[!t]
    \centering
    \includegraphics[width=0.9\columnwidth]{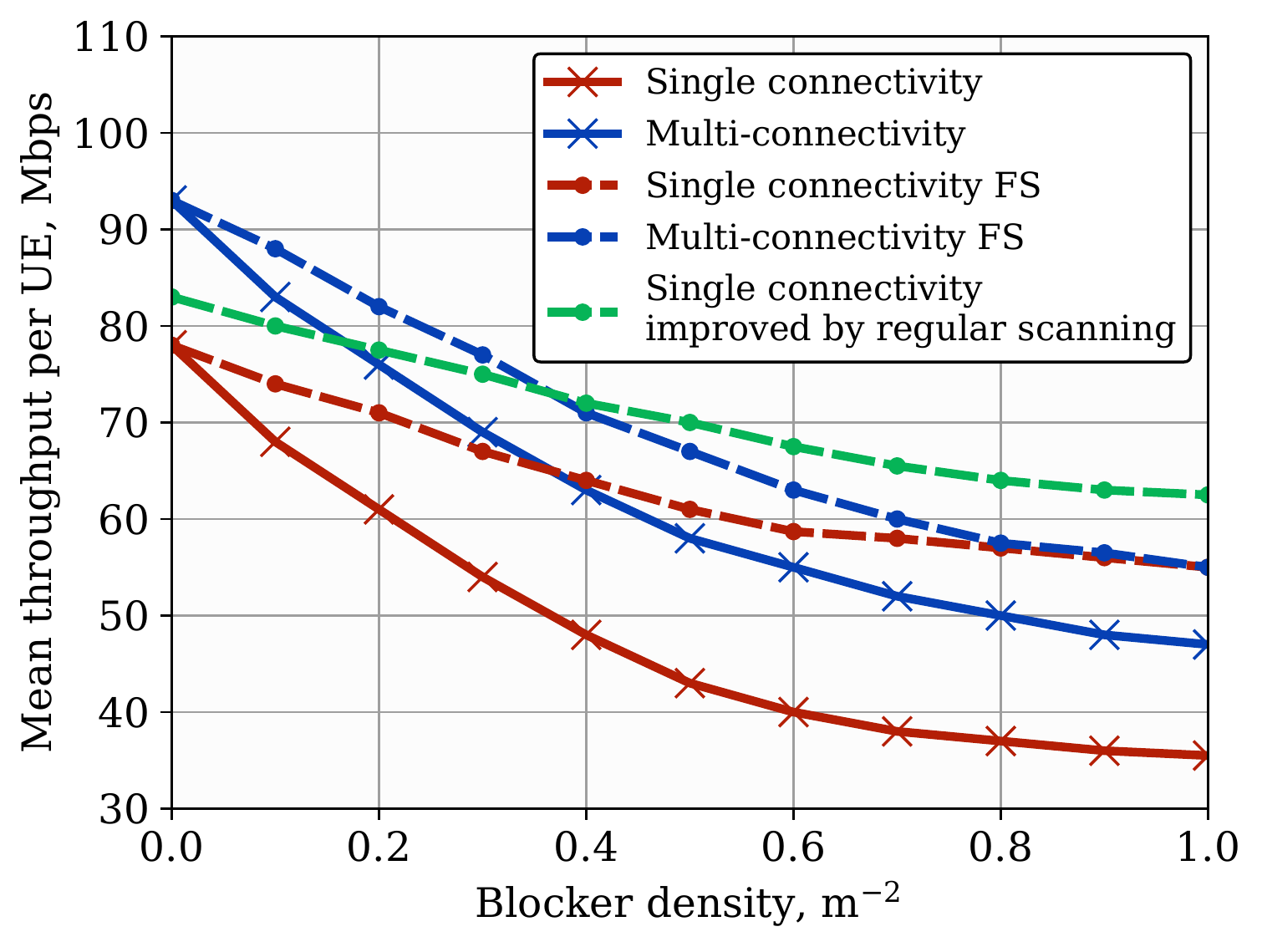}
    \caption{Mean UE throughput as a function of blocker density with and without multi-connectivity capabilities.}
    \label{fig:multi_con}
    \vspace{-4mm}
\end{figure}

Another mechanism that we consider is multi-connectivity. To this aim, Fig. \ref{fig:multi_con} shows the mean UE throughput for UL and DL session arrival rates of $0.5$ sessions per second as a function of the blocker density. In this setup, PF slot formatting, association scheme based on the RSRP, and single beam operation of the DgNB are utilized.


Analyzing the results shown in Fig. \ref{fig:multi_con}, one may observe that the conventional multi-connectivity option outperforms single connectivity operation by approximately $15$\% across the considered range of blocker densities. The main reason is that the former allows exploiting more resources available at the IAB nodes and the DgNB. However, with the FS capabilities, single connectivity outperforms the conventional multi-connectivity scheme. 

The rationale behind the obtained results is that the use of FS capabilities not only allows to efficiently combat the effect of blockage but also to evenly distribute the network load across the IAB nodes and the DgNB. By further utilizing regular scanning, one may improve the load distribution even further. This conclusion is important for the UE energy conservation as single-connectivity with FS and regular scanning do not require the support of two active links. \textcolor{MajorRevision}{At the same time, this configuration shows higher UE throughput across the considered range of parameters. }


The reported behavior also emphasizes the importance of balanced traffic distribution in the IAB deployments. Particularly, the choice of an uncongested route to the DgNB becomes more essential than the choice of a link having slightly better channel conditions. In this context, we specifically emphasize the aforementioned \textcolor{MajorRevision}{\enquote{self-balancing}} behavior of the considered  mmWave IAB system, where one may not require any further mechanisms to ensure equal load distribution. 



\subsection{Multi-Beaming at DgNB and/or IAB nodes}

\begin{figure}[!b]
    \vspace{-4mm}
    \centering
    \includegraphics[width=0.9\columnwidth]{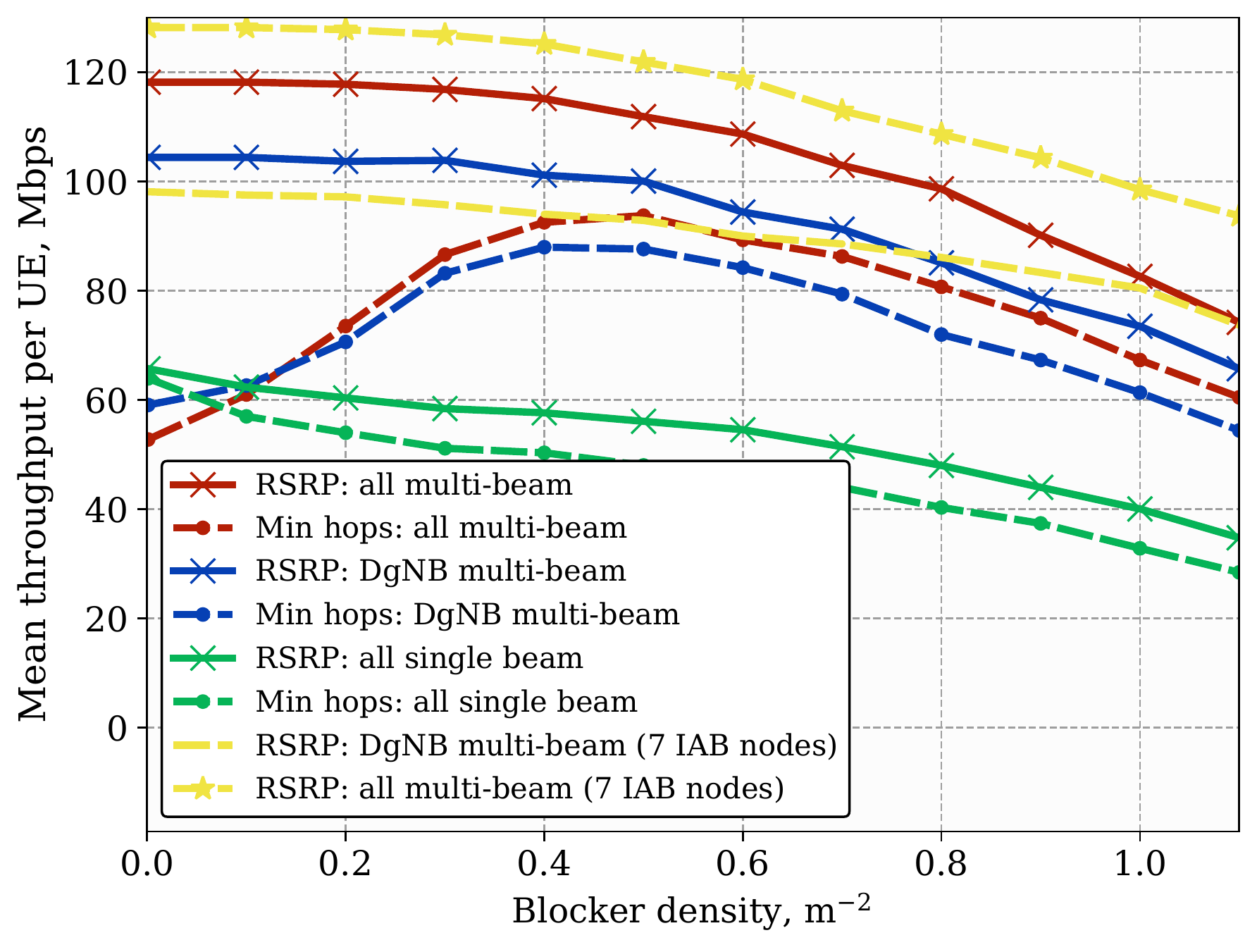}
    \caption{\textcolor{MajorRevision}{Mean UE throughput as a function of blocker density for different association schemes with and without multi-beam functionality.}}
    \label{fig:multi_beam}
    \vspace{-0mm}
\end{figure}

Another functionality that we address is multi-beam DgNB and IAB node operation as illustrated in Fig. \ref{fig:multi_beam}. The latter shows the mean UE throughput as a function of blocker density with the UL and DL session arrival intensities of $0.5$. We consider the option of single connectivity with FS and regular scanning capabilities that showed the best results previously. 
Multi-beaming at the DgNB improves the UE performance by around $50-70$\% depending on the blocker intensities. However, introducing multi-beam support at the IAB nodes improves it further by only $10-15$\%. \textcolor{MajorRevision}{The reason is that changing the DgNB configuration to multi-beam allows overcoming the backhaul-limited regime that occurs in the single-beam mode.} 



By comparing the performance of route selection schemes also indicated in Fig. \ref{fig:multi_beam}, one may observe that the throughput attained by the best RSRP option is higher compared to the minimum number of hops scheme. For the latter scheme, the throughput initially increases. The rationale is that as the blocker density grows, blockage of direct links with the DgNB leads to choosing other routes that have multiple hops, which benefits the system performance. That is, increased blocker density leads to better topology under the minimum number of hops strategy. \textcolor{MajorRevision}{The higher number of IAB nodes provides an additional pool of resources in the multi-beam case. Without multi-beam functionality, this leads to the increased delay.}




\section{Challenges and Future Work}\label{sect:challenges}

The introduction of multi-hop operation in 5G IAB architecture naturally requires semi-distributed operation in cellular systems that historically relied upon fully centralized control. This opens up several unique research questions that need to be carefully addressed.


\subsubsection{RA and Topology Organization}

The use of multi-hop communications brings challenges related to efficient RA. Centralized RA involves additional latency related to the delivery of decisions to remote IAB nodes. Hence, the overall RA in IAB systems needs to be performed in a semi-distributed manner. 
The matter of RA in IAB systems cannot be considered separately from topology optimization. Therefore, RA needs to account for multiple paths available and may include traffic splitting functionality. 
Operating in a full-duplex mode, this can be formalized as a conventional network flow problem \cite{pioro2004routing}. However, the half-duplex regime adds further constraints to the network communications patterns.




\subsubsection{Distributed Packet Scheduling}

MAC scheduling algorithms pose another challenge to the IAB system. In our simulation, the equal time-sharing scheduler is utilized as the focus is to demonstrate the potential of dynamic slot formatting. However, the utilization of another scheduler may provide different quantitative results. Moreover, the related overhead cannot be disregarded. For example, RRC signaling is necessary to configure a slot format as discussed further. In addition, a guard interval is required for a transceiver to switch between the UL and DL directions. It is worth noting that the slot format should be configured in the presence of the half-duplex constraint. 


\subsubsection{Multi-Hop and Multi-Connectivity}

In real deployments, the density of blockers varies on, e.g. time of the day; thus, it may be beneficial to enable adaptability of the system. In addition, multi-connectivity solutions require synchronization of the scheduling and ARQ mechanisms. While the utilization of different paths provides channel diversity, it also results in packet delay variability. Therefore, packets arriving out of order may create bufferbloat if no reordering algorithm is applied. When switching between the available links, one should take into account the delay due to the data collecting and beam sweeping procedures.



\subsubsection{Multi-beam Operation}

This operation raises research questions related to the optimal number of beams required to achieve the desired balance between access and backhaul limited regimes.
Moreover, the use of digital beamforming at the IAB nodes depends on the cost-efficiency trade-off that has not been deeply addressed so far. On the one hand, it allows enhancing the backhaul capacity at IAB interfaces. However, it lowers the power of each beam and may also create additional interference for individual UEs. Similar to the use of this functionality at the DgNB, one needs to determine the optimal number of beams required to fully utilize the available resources.


\subsubsection{Signaling Overheads}

Operating in a semi-decentralized regime, IAB systems naturally face challenges related to signaling. \textcolor{MajorRevision}{On the other hand, beam management signaling overhead will be reduced in 3GPP Release 17}. The type of information that needs to be exchanged between UE/IAB nodes to DgNB to make decisions on RA and topology maintenance is not specified in 3GPP documents. Potentially, the information provided to the DgNB by the IAB nodes may include buffer states of UEs, their capabilities, QoS, current resource utilization of backhaul and access links, etc. However, due to the limited capacity of control channels, propagation, and buffering delays along multi-hop routes, this information is limited.

\section{Conclusions}



%

3GPP IAB architecture promises to bring a cost-efficient means of densifying the 5G cellular deployments by providing both coverage extension and capacity boost at the air interface.
Our results demonstrated that for a cost-efficient deployment of 3GPP IAB systems, where multi-beam and multi-connectivity functionalities are not utilized, the throughput gain from enabling dynamic slot formatting over multi-hop topologies is notable and reaches $10 - 30$ \%. By employing multi-connectivity operation with advanced link switching mechanisms, the system can reach further capacity gains of $10-40$ \% depending on the density of blockers. Furthermore, the use of dynamic link selection strategies not only efficiently mitigates the impact of dynamic blockage but also equalizes the load across the IAB nodes and DgNB, thus resulting in more efficient use of the available resources. Finally, multi-beam operation yields better performance irrespective of the choice of other system parameters. However, most of these gains stem from enabling this advanced functionality at the DgNB side. 

\bibliographystyle{ieeetr}
\bibliography{lib}
\balance

\textbf{Yekaterina~Sadovaya} is with the Unit of Electrical Engineering, Tampere University, Finland. Her research interests include the development of future cellular systems and millimeter-wave communication.
	
\textbf{Dmitri~Moltchanov} is University Lecturer at TAU. He received his Ph.D. degree from TUT. His current research interests include research and development of 5G/5G+ systems, URLLC, IoT, and V2V/V2X systems.

\textbf{Wei~Mao} received his Ph.D. degree from California Institute of Technology. In 2017 he joined Intel Corporation, Santa Clara, CA, as a research scientist. 

\textbf{Oner~Orhan} received the Ph.D. degree from New York University Tandon School of Engineering. He currently works at Intel labs as an AI/ML research scientist. 

\textbf{Shu-ping~Yeh} is a Senior Research Scientist in the Wireless System Research Lab at Intel. Dr. Yeh has over 10 years of research and development experience in wireless industry. She received her M.S. and Ph.D. in Electrical Engineering from Stanford University.

\textbf{Hosein~Nikopour} is a Senior Research Scientist and Manger in the Intel Labs. Prior to Intel, he was with Huawei Canada. Hosein is coinventor of more than 70 patents and several publications with more than 4000 citations.
    
\textbf{Shilpa~Talwar} is an Intel Fellow and director of wireless multi-communication systems with Intel Labs. She is co-editor of book on 5G “Towards 5G: Applications, requirements and candidate technologies.” Shilpa graduated with PhD from Stanford University, and is the author of 70 publications and 60 patents.

\textbf{Sergey~Andreev} is an associate professor of communications engineering and Academy Research Fellow at Tampere University, Finland. He has (co-)authored more than 200 published research works. 

\end{document}